\documentclass[aps,preprint,psfig,showpacs]{revtex4}
\usepackage{amsfonts}
\usepackage{amsmath}
\usepackage{graphicx}
\usepackage{bm}
\usepackage{amssymb}
\usepackage{times}
\usepackage{dcolumn}

\makeatletter

\begin{document}

\title{Spin transfer in a ferromagnet-quantum dot and tunnel barrier coupled
Aharonov-Bohm ring system with Rashba spin-orbit interactions}
\author{Xi Chen$^{1,2}$, Qing-Rong Zheng$^{1}$ and Gang Su$^{1,}$}
\email[Author to whom correspondence should be addressed. ]{ Email:
gsu@gucas.ac.cn}
\affiliation{$^{1}$College of Physical Sciences, Graduate University of Chinese Academy
of Sciences, P.O. Box 4588, Beijing 100049, China\\
$^{2}$Institut de Physique et de Chimie des Mat\'{e}riaux de Strasbourg
(IPCMS), UMR 7504 CNRS-ULP, 23 rue de loess, 67034 Strasbourg, France}

\begin{abstract}
The spin transfer effect in ferromagnet-quantum dot (insulator)-ferromagnet
Aharonov-Bohm (AB) ring system with Rashba spin-orbit (SO) interactions is
investigated by means of Keldysh nonequilibrium Green function method. It is
found that both the magnitude and direction of the spin transfer torque
(STT) acting on the right ferromagnet electrode can be effectively
controlled by changing the magnetic flux threading the AB ring or the gate
voltage on the quantum dot. The STT can be greatly augmented by matching a
proper magnetic flux and an SO interaction at a cost of low electrical
current. The STT, electrical current, and spin current are uncovered to
oscillate with the magnetic flux. The present results are expected to be
useful for information storage in nanospintronics.
\end{abstract}

\maketitle

\section{Introduction}

The spin transfer effect (STE) states that when the spin-polarized electrons
flow from one ferromagnet (FM) layer into another FM layer with
magnetization aligned by a relative angle, they may transfer transverse spin
angular momenta to the local spins of the second FM layer, thereby exerting
a torque on the magnetic moments that is usually coined as the spin transfer
torque (STT). This important phenomenon was predicted independently by
Berger and Slonczeski \cite{J.C,Berger} in 1996 and soon confirmed by
experiments. Because the STE can be utilized to switch the magnetic state of
the free FM layer in a magnetic tunneling junction (MTJ) or a spin valve by
applying an electrical current instead of a magnetic field, it may be even
more useful in writing heads for magnetic random access memory (MRAM) or
hard disk drivers than the conventional tunnel magnetoresistance (TMR) and
giant magnetoresistance (GMR) effects. In view of the potentially wide
applications in nanospintronic devices, a number of works on the STE have
been done for different systems both theoretically and experimentally \cite%
{Zhu,Mu1,Mu2,Ioannis,Z.Z.Sun,G.D,Levy,J.A,Z.Li,K.Xia,G.D2,S.Zhang,M.D,K.Ando}%
.

On the other hand, the quantum dot (QD) has received much attention in the
past decades, and a lot of advances have been made in this particular field
(e.g. Refs. \cite{I.Z,Hanson,J.Mar,Xi,Xi2}). For a semiconductor QD, as the
spin-orbit (SO) interaction is usually not negligible, some interesting
phenomena related to the SO interactions, such as the bias-controllable
intrinsic spin polarization in a QD \cite{Sun2} and the interplay of Fano
and Rashba effect \cite{JPCM}, can be observed. Almost twenty years ago,
Datta and Das predicted a spin transistor based on the Rashba SO \cite{DD},
showing that the SO interactions may be important in the semiconductor
spintronics. However, the effect of the Rashba SO interaction on the STE is
still sparsely studied. In this paper, we shall take the FM-QD (insulator,
I)-FM Aharonov-Bohm (AB) ring system as an example to investigate how both
the SO interaction and the magnetic flux affect the spin-dependent
properties of the system by means of the nonequilibrium Green function
method. We have found that the magnitude and direction of the STT can be
easily controlled by changing the gate voltage $V_{g}$ on the QD or the
magnetic flux $\phi $ through the ring if both the SO and electron-electron
(e-e) interactions in the QD are considered, which might be useful in
information storage.

The other parts of this paper are organized as follows. In Sec. II, a model
is proposed, and the relevant Green functions are obtained in terms of the
nonequilibrium Green function method. In Sec. III, the spin-dependent
properties of STT in the system under interest are numerically investigated,
and some discussions are presented. Finally, a brief summary is given in
Sec. IV.

\section{MODEL AND METHOD}

The system under interest is depicted in Fig. 1. Two FM leads spreading
along the $z$ axis are weakly coupled to an insulating (I) barrier and a
semiconducting QD, forming an AB ring. The left (L) FM electrode with the
magnetization along the $z$ axis is applied by a bias voltage $-V/2$, while
the right (R) electrode with the magnetization along the $z^{\prime }$ axis
that deviates by an angle $\theta $ from the $z$ axis is applied by a bias
voltage $V/2$. Assume that the QD is made of a two-dimensional electron gas
in which the electrons are strongly confined in the $y$ direction by a
potential $V(y)$. Due to $dV/dy\gg dV/dx$ and $dV/dz$, we have $\nabla V(%
\overrightarrow{y})\approx \widehat{y}(dV/dy)$, where $\widehat{y}$ is the
unit vector along the $y$ axis. If $V(y)$ is asymmetric to $y=0$, both
Rashba SO and e-e interactions on the QD should be considered. Since the
electronic transport of the device along the $z$ axis is much more dominant
than that along other two dimensions, the device under interest can be
treated as a quasi one-dimensional system. Sun \textit{et al}. \cite{Sun1}\
have carefully analyzed the SO Rashba interaction and found that (i) the
Rashba SO interaction can be separated into two parts, $H_{R_{1}}$ and $%
H_{R_{2}}$, namely
\begin{eqnarray}
H_{so} &=&\frac{\widehat{y}}{2\hbar }\cdot \lbrack \alpha (x)(\widehat{%
\sigma }\times \widehat{p})+(\widehat{\sigma }\times \widehat{p})\alpha
(x)]=H_{R_{1}}+H_{R_{2}}, \\
H_{R_{1}} &=&\frac{1}{2\hbar }[\alpha (x)\sigma _{z}p_{x}+\sigma
_{z}p_{x}\alpha (x)], \\
H_{R_{2}} &=&-\frac{\alpha (x)\sigma _{x}p_{z}}{\hbar };
\end{eqnarray}%
(ii) by choosing a suitable unitary transformation, $H_{R_{1}}$ can
give rise to a spin-dependent phase factor in the tunneling matrix
element between the leads and the QD, while Eq. (3) can be written
in the second-quantization form as \cite{Sun1}: $H_{R_{2}}=\sum%
\limits_{mn}(t_{mn}^{so}d_{m\downarrow }^{+}d_{n\uparrow }+h.c.)$,
that causes a spin-flip term with strength $t_{mn}^{so}$ in the QD,
where $m$ and $n$ are quantum numbers for the eigenstates of
electrons in QD; (iii) since the
time-reversal invariance is maintained by the Rashba SO interaction, $%
t_{mn}^{so}=-t_{nm}^{so}$ and $t_{nn}^{so}=0$, which suggests that the
spin-flip scatterings only occur between different levels in the QD. In the
present work, for simplicity, we shall consider the case with a single-level
QD as in some previous works \cite{Sun1,JPCM}, where no interlevel spin-flip
scattering happens in the QD. Thus, $H_{R_{2}}$ equals to zero. Suppose that
$\alpha (x)$ is independent of the coordinates in the scattering region, and
a magnetic flux penetrates into the AB ring. The Hamiltonian of the present
system is given by

\begin{equation}
H=H_{QD}+H_{\beta }+H_{T},
\end{equation}

\begin{equation}
H_{QD}=\sum\limits_{\sigma }\varepsilon _{d}d_{\sigma }^{+}d_{\sigma
}+Un_{\uparrow }n_{\downarrow },
\end{equation}

\begin{equation}
H_{\beta }=\sum\limits_{\beta k,\sigma }\varepsilon _{\beta k\sigma
}a_{\beta k\sigma }^{+}a_{\beta k\sigma },
\end{equation}

\begin{eqnarray}
H_{T} &=&\sum\limits_{k,\sigma }[t_{Rd}(\cos \frac{\theta _{\beta }}{2}%
a_{Rk\sigma }^{+}-\sigma \sin \frac{\theta _{\beta }}{2}a_{Rk\overline{%
\sigma }}^{+})\times e^{-i\sigma \gamma }e^{i\phi }d_{\sigma }+h.c.]  \notag
\\
&&+\sum\limits_{k,\sigma }[t_{Ld}a_{Lk\sigma }^{+}d_{\sigma }+h.c.]  \notag
\\
&&+\sum\limits_{k,\sigma }[t_{LR}(\cos \frac{\theta _{\beta }}{2}a_{Rk\sigma
}^{+}-\sigma \sin \frac{\theta _{\beta }}{2}a_{Rk\overline{\sigma }%
}^{+})a_{Lk\sigma }+h.c.],
\end{eqnarray}%
where $a_{\beta k\sigma }$ and $d_{\sigma }$ are annihilation operators of
electrons with momentum $k$ and spin $\sigma $ $(=\uparrow ,\downarrow )$ in
the $\beta $ $(=L,R)$ electrode and in the QD, respectively, $\varepsilon
_{\beta k\sigma }=\varepsilon _{k}+\sigma {}M_{\beta }-eV_{\beta }$ is the
single-electron energy for the wave vector $k$ with the molecular field $%
M_{\beta }$ in the electrode $\beta $, $\varepsilon _{d}$ is the
single-electron energy in the QD, $U$ represents the on-site Coulomb
interaction between electrons in the QD, $t_{\beta d}$ is the tunneling
matrix element of electrons between the $\beta $ electrode and the QD, $%
t_{LR}$ is the tunneling matrix element of electrons between $L$ and $R$
electrodes through the insulating barrier, $n_{\sigma }=c_{\sigma
}^{+}c_{\sigma }$, and $\gamma =$ $k_{R}d$ with $k_{R}\equiv \alpha m^{\ast
}/\hbar ^{2}$, $\alpha =\langle \Psi (y)\left\vert (d/dy)V(y)\right\vert
\Psi (y)\rangle $, $m^{\ast }$ the effective mass of electrons and $d$ the
thickness of the middle region. The magnetic flux $\Phi $ threading the AB
ring is related to the phase factor by $\phi =2\pi \Phi /\Phi _{0}$, where $%
\Phi _{0}$ is the flux quantum. It should be noted that the magnetic flux
threading the AB ring generally includes two contributions, one generated by
the FM leads that may be small and constant, and the other from the external
magnetic field that can be varied to adjust the phase factor $\phi $.
\begin{figure}[tbp]
\vspace{-1cm} \center{\includegraphics[width=0.6\linewidth,clip]{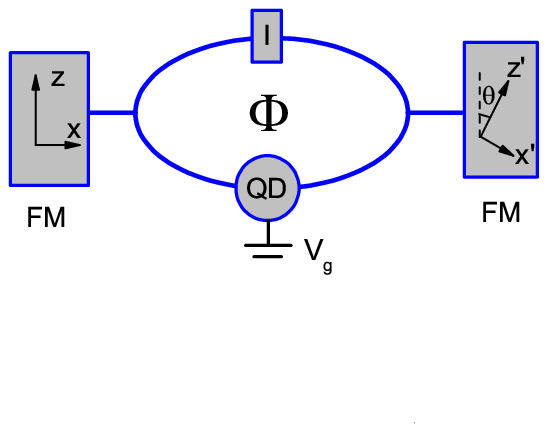}}
\vspace{-2.0cm}
\caption{A schematic layout of FM-QD(I)-FM AB ring system. The electrons
flow towards the $x$ axis.}
\end{figure}

The transverse component of the total spin in the right FM lead can be
written as \cite{Zhu}
\begin{eqnarray}
S &=&\frac{\hbar }{2}\sum\limits_{k}\left( a_{Rk\uparrow
}^{+},a_{Rk\downarrow }^{+}\right) \hat{\sigma}_{x}\left(
\begin{array}{c}
a_{Rk\uparrow } \\
a_{Rk\downarrow }%
\end{array}%
\right)  \notag \\
&=&\frac{\hbar }{2}\sum\limits_{k}(a_{Rk\uparrow }^{+}a_{Rk\downarrow
}+a_{Rk\downarrow }^{+}a_{Rk\uparrow }),
\end{eqnarray}%
where $S$ is written in the $x^{\prime }y^{\prime }z^{\prime }$ coordinate
frame. The spin torque, namely, the time evolution rate of the transverse
component of the total spin of the right FM lead, can be obtained by $%
\partial S/\partial t=\frac{i}{\hbar }\langle \lbrack H,S]\rangle $.
According to Refs. \cite{X.W1,X.W2,Zhu}, the right FM layer gains two types
of torques: one is the equilibrium torque caused by the spin-dependent
potential, and another is from the tunneling of electrons that is in what we
are interested. After cautiously separating the current-induced torque from
the equilibrium one, the STT is given by
\begin{eqnarray}
\tau &=&-\Re e\{t_{Rd}e^{i\gamma }e^{i\phi }\cos \frac{\theta }{2}%
G_{d\downarrow R\uparrow }^{<}(t,t)-t_{Rd}e^{-i\gamma }e^{i\phi }\sin \frac{%
\theta }{2}G_{d\uparrow R\uparrow }^{<}(t,t)  \notag \\
&&+t_{RL}\cos \frac{\theta }{2}G_{L\downarrow R\uparrow
}^{<}(t,t)-t_{RL}\sin \frac{\theta }{2}G_{L\uparrow R\uparrow }^{<}(t,t)
\notag \\
&&+t_{Rd}e^{-i\gamma }e^{i\phi }\cos \frac{\theta }{2}G_{d\uparrow
R\downarrow }^{<}(t,t)+t_{Rd}e^{i\gamma }e^{i\phi }\sin \frac{\theta }{2}%
G_{d\downarrow R\downarrow }^{<}(t,t)  \notag \\
&&+t_{RL}\cos \frac{\theta }{2}G_{L\uparrow R\downarrow
}^{<}(t,t)+t_{RL}\sin \frac{\theta }{2}G_{L\downarrow R\downarrow
}^{<}(t,t)\}.
\end{eqnarray}

From Eq. (9), it is clear that the current-induced STT can be obtained as
long as we get the lesser Green functions $G^{<}$. In what follows we shall
use Keldysh's nonequilibrium Green function technique to determine all
lesser Green functions \cite{Jauho}. These functions are closely related to
the retarded Green functions defined by

\begin{eqnarray*}
G_{\beta \sigma \gamma \sigma ^{\prime }}^{r}(t,t^{\prime }) &=&-i\theta
(t-t^{\prime })\langle \{\sum\limits_{k^{^{\prime }}}a_{\beta k^{^{\prime
}}\sigma }(t),\sum\limits_{k}a_{\gamma k\sigma ^{\prime }}^{+}(t^{\prime
})\}\rangle , \\
G_{\beta \sigma d\sigma ^{\prime }}^{r}(t,t^{\prime }) &=&-i\theta
(t-t^{\prime })\langle \{\sum\limits_{k}a_{\beta k\sigma }(t),d_{\sigma
^{\prime }}^{+}(t^{\prime })\}\rangle , \\
G_{d\sigma d\sigma ^{^{\prime }}}^{r}(t,t^{\prime }) &=&-i\theta
(t-t^{\prime })\langle \{d_{\sigma }(t),d_{\sigma ^{\prime }}^{+}(t^{\prime
})\}\rangle ,
\end{eqnarray*}%
where $\{A,B\}$ denotes the anticommutation relations, and $\langle A\rangle
$ stands for the thermal average. By using the equation of motion, the
retarded Green functions can be obtained by Dyson equation $%
G^{r}=g^{r}+g^{r}\Sigma ^{r}G^{r}$, where $g^{r}$ is the retarded Green
function for decoupled systems, and $\Sigma ^{r}$ is the self-energy of
electrons. To obtain $G^{r}$, the decoupling approximations similar to those
in Refs. \cite{Mu2,Xi3,Jauho} for the equations of motion of Green functions
should be made. As the associated equations for Green functions are quite
lengthy, we shall not repeat them here for conciseness.

The lesser Green function $G^{<}$ can be calculated straightforwardly from
the Keldysh equation%
\begin{eqnarray}
G^{<} &=&(1+G^{r}\Sigma ^{r})g^{<}(1+\Sigma ^{a}G^{a})+G^{r}\Sigma ^{<}G^{a}
\notag \\
&=&G^{r}g^{r-1}g^{<}g^{a-1}G^{a}+G^{r}\Sigma ^{<}G^{a}.
\end{eqnarray}

In the present case, $\Sigma ^{<}=0$, and $g^{r-1}g^{<}g^{a-1}$ is diagonal

\begin{equation}
g^{r-1}g^{<}g^{a-1}=\left(
\begin{array}{cccccc}
2if_{L}(\varepsilon )/\pi \rho _{L\uparrow } &  &  &  &  &  \\
& 2if_{L}(\varepsilon )/\pi \rho _{L\downarrow } &  &  &  &  \\
&  & 2if_{R}(\varepsilon )/\pi \rho _{R\uparrow } &  &  &  \\
&  &  & 2if_{R}(\varepsilon )/\pi \rho _{R\downarrow } &  &  \\
&  &  &  & 0 &  \\
&  &  &  &  & 0%
\end{array}%
\right) .
\end{equation}%
The electrical current is given by

\begin{eqnarray}
I &=&I_{\uparrow }+I_{\downarrow }, \\
I_{\uparrow } &=&-\Re e\{t_{Rd}e^{-i\gamma }e^{i\phi }\cos \frac{\theta }{2}%
G_{d\uparrow R\uparrow }^{<}(t,t)+t_{Rd}e^{i\gamma }e^{i\phi }\sin \frac{%
\theta }{2}G_{d\downarrow R\uparrow }^{<}(t,t)  \notag \\
&&+t_{RL}\cos \frac{\theta }{2}G_{L\uparrow R\uparrow }^{<}(t,t)+t_{RL}\sin
\frac{\theta }{2}G_{L\downarrow R\uparrow }^{<}(t,t)\}, \\
I_{\downarrow } &=&-\Re e\{t_{Rd}e^{i\gamma }e^{i\phi }\cos \frac{\theta }{2}%
G_{d\downarrow R\downarrow }^{<}(t,t)-t_{Rd}e^{-i\gamma }e^{i\phi }\sin
\frac{\theta }{2}G_{d\uparrow R\downarrow }^{<}(t,t)  \notag \\
&&+t_{RL}\cos \frac{\theta }{2}G_{L\downarrow R\downarrow
}^{<}(t,t)-t_{RL}\sin \frac{\theta }{2}G_{L\uparrow R\downarrow }^{<}(t,t)\}.
\end{eqnarray}%
The spin current is defined by a difference between the electrical currents
of spin up and down,
\begin{equation}
I_{s}=I_{\uparrow }-I_{\downarrow }.
\end{equation}%
To get the physical quantities of interest, the above-mentioned equations
will be solved numerically in a self-consistent manner.

\section{RESULTS AND DISCUSSIONS}

It has been shown that when the incident electrical current is larger than a
critical value, the STT can switch the direction of the magnetization of the
free FM layer clockwise or anticlockwise depending on the direction of the
incident electrical current \cite{J.Z,J.Z.Sun,Xi3}. In the present case, the
positive STT tends to push the spins in the right FM electrode aligning
antiparallel with the magnetization of the left FM electrode, while the
negative STT may cause a reverse orientation of the magnetization in the
free FM layer. In order to properly incorporate the STE into a
functionalized spintronic device, both the direction and magnitude of the
STT should be taken into account. For simplicity, in the following parts we
will assume that in most cases the left and right FM electrodes have the
same spin polarization $P_{L}=P_{R}=P=0.5$, and the angle $\theta $ between $%
z$ and $z^{\prime }$ axes is $\pi /3$ throughout the paper unless specified.
We take $I_{0}=\frac{e\Gamma _{0}}{\hbar }$ and $\Gamma _{0}=\Gamma
_{L(R)\uparrow }(P=0)=\Gamma _{L(R)\downarrow }(P=0)$ as scales for the
electrical and spin currents as well as the STT and energy, respectively,
where $\Gamma _{\alpha \sigma }(\varepsilon )=2\pi \sum\nolimits_{k_{\alpha
}}\left\vert t_{\alpha d}\right\vert ^{2}\delta (\varepsilon -\varepsilon
_{k_{\alpha }})$. In accordance with Refs. \cite{F.M,T.M,D.G}, we assume
that the Rashba SO interaction constant is $\alpha \sim 3\times 10^{-11}eVm$%
, $k_{R}=m^{\ast }\alpha /\hbar ^{2}\approx 0.015/nm$ for $m^{\ast
}=0.036m_{e}$, the typical length of QD is $100$ $nm$, $U=5\Gamma _{0}$, and
$\gamma $ can be $\pi /2$ or larger.

\begin{figure}[tbp]
\includegraphics[width=0.6\linewidth,clip]{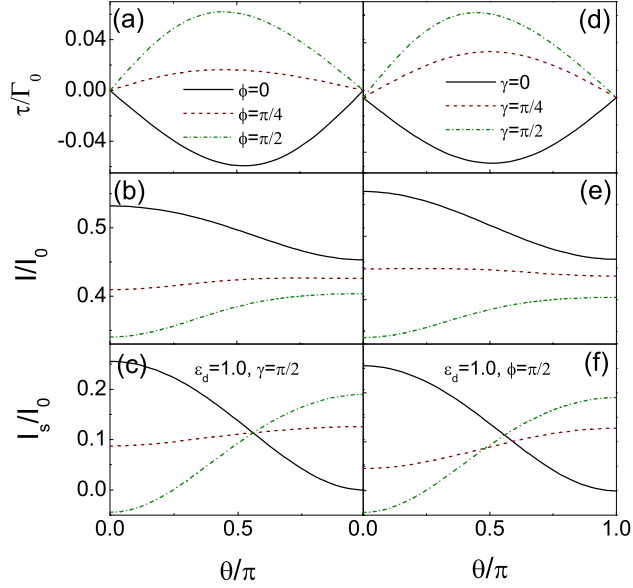}
\caption{(Color Online) The $\protect\theta $ dependence of (a) the spin
transfer torque $\protect\tau $, (b) the electrical current $I$, and (c) the
spin current $I_{s}$ for different magnetic flux $\protect\phi $. The $%
\protect\theta $ dependence of (d) the spin transfer torque $\protect\tau $,
(e) the electrical current $I$, and (f) the spin current $I_{s}$ for
different Rashba SO interaction $\protect\gamma $.}
\end{figure}

As the STE only exists in the noncollinear case, in contrast to previous
works where the electrical and spin currents were discussed only in
collinear cases ($\theta =0$ or $\pi $) (e.g. \cite{Sun1,JPCM}), let us
first look at the angular dependences of the STT, the electrical current and
spin current for different magnetic flux $\phi $ and Rashba SO interaction $%
\gamma $. The results are given in Fig. 2, where $\gamma =\pi /2$ and $%
\varepsilon _{d}=1$ in Figs. 2(a)-(c), and $\phi =\pi /2$ and $\varepsilon
_{d}=1$ in Figs. 2(d)-(f). It can be observed that the STT has a sine-like
relationship with the relative angle $\theta $, and the direction and
magnitude of the STT are clearly influenced by both the magnetic flux $\phi $
and Rashba SO interaction $\gamma $. In the absence of either $\phi $ or $%
\gamma $, the STT remains negative (anticlockwise). For the simultaneous
presence of $\phi $ and $\gamma $ (greater than $\pi /4$), the STT becomes
positive (clockwise). This fact reminds us that we may apply the magnetic
flux to change the direction of the STT, thereby being capable of
manipulating the magnetic state of the free FM layer, which might be useful
for information storage and for designing the memory element. The electrical
current $I$ decreases with increasing $\theta $ in the absence of either $%
\phi $ or $\gamma $, indicating a spin-valve effect, while it increases with
$\theta $ in the presence of both $\phi $ and $\gamma $ (greater than $\pi
/4 $), giving an anti-spin-valve effect [e.g. $\phi $ or $\gamma =\pi /2$ in
Figs. 2(b) and (e)]. This property differs obviously from the conventional
FM-I-FM or FM-QD-FM systems without considering the SO interactions where
the electrical current always decreases with increasing $\theta $. The spin
current shows a feature similar to the electrical current. From these
calculated results presented in Fig. 2, we can find that for a given Rashba
SO interaction $\gamma $ (magnetic flux $\phi $), the angular dependent STT,
electrical current and spin current exhibit distinct behaviors for different
magnetic flux (Rashba SO interaction).

\begin{figure}[tbp]
\includegraphics[width=0.6\linewidth,clip]{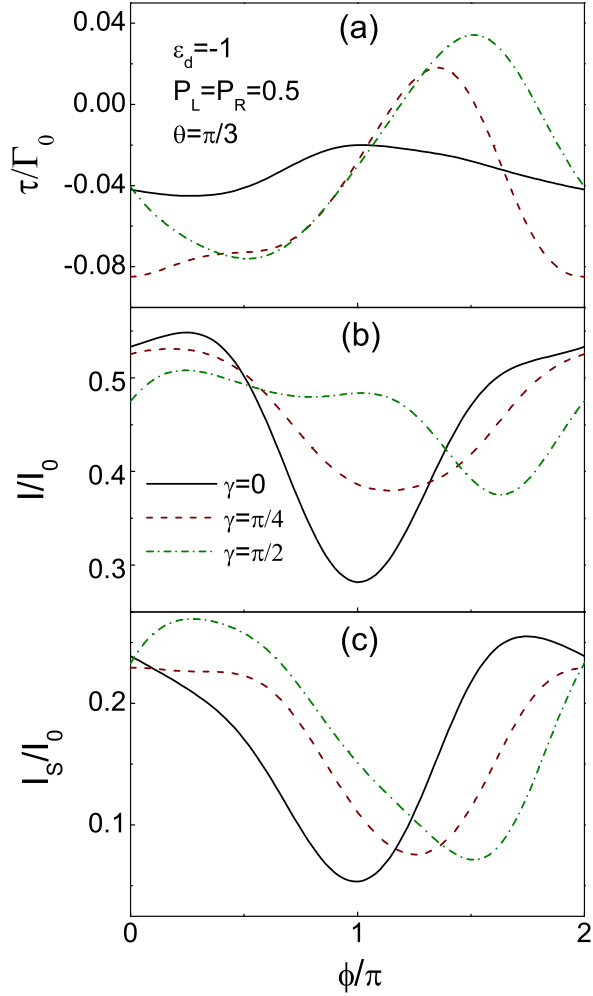}
\caption{(Color Online) The magnetic flux dependence of (a) spin transfer
torque $\protect\tau $, (b) electrical current $I$, and (c) spin current $%
I_{s}$ for different Rashba SO interaction $\protect\gamma $, where $\protect%
\varepsilon _{d}=-1$, and $\protect\theta =\protect\pi /3$.}
\end{figure}

Figure 3 shows the magnetic flux $\phi $ dependence of the STT, electrical
current and spin current for different Rashba SO interactions. We can see
that with increasing $\phi $, the STT, electrical current and spin current
oscillate differently for various Rashba SO interactions $\gamma $. The
larger the SO interaction $\gamma $ is, the more complex the oscillations
are. For the QD energy level $\varepsilon _{d}=-1$ and $\gamma =\pi /2$, the
STT shows a maximum around $\phi =3\pi /2$, while the electrical current
exhibits minima around the same $\phi $. Therefore, we may be able to use a
lower current to change the magnetic state of the free FM by adjusting the
magnetic flux penetrating into the AB ring. It is favorable for the spintronic devices, because a larger current may cause more heating, while the heating should be reduced as small as possible for better functions of the device. In addition, it can be found that
the STT is closely related to the spin current, as the dips and peaks of
Figs. 3(a) and (c) appear almost at the same positions.

\begin{figure}[tbp]
\includegraphics[width=0.6\linewidth,clip]{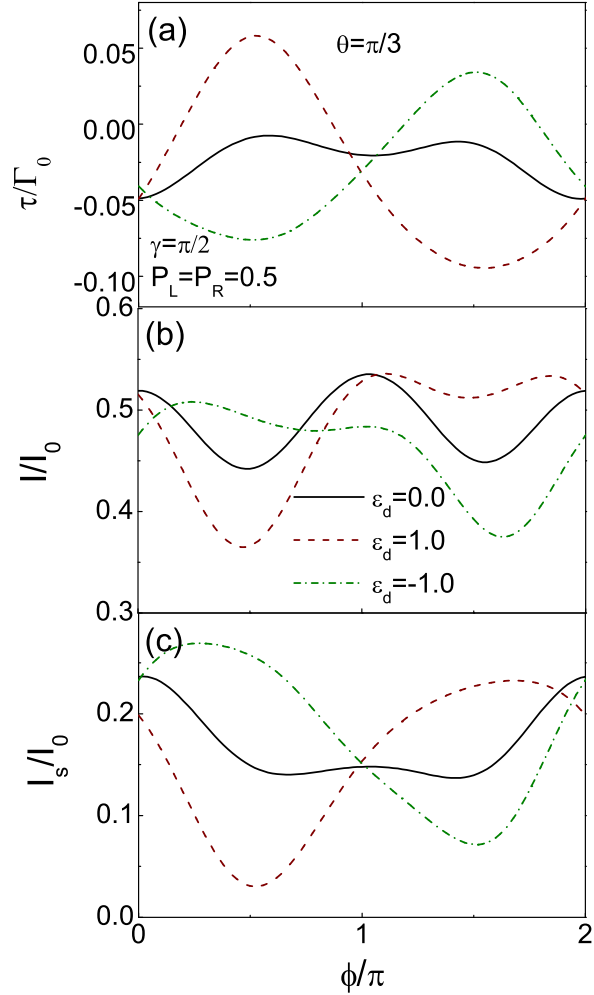}
\caption{(Color Online) The magnetic flux dependence of (a) the spin
transfer torque $\protect\tau $, (b) the electrical current $I$, and (c) the
spin current $I_{s}$ for different energy levels $\protect\varepsilon _{d}$,
where $\protect\theta =\protect\pi /3$, and $\protect\gamma =\protect\pi /2$%
. }
\end{figure}

The energy level $\varepsilon _{d}$ of electrons in the QD, that offers
resonant tunneling channels for spin-polarized electrons from the left FM
electrode to the right FM one, has also effects on the magnetic flux $\phi $
dependence of the STT, electrical current and spin current. The results are
presented in Fig. 4 for $\gamma =\pi /2$. It is unclosed that for different $%
\varepsilon _{d}$, $\tau $, $I$ and $I_{s}$ exhibit different features, and
oscillate with $\phi $ in general. When $\varepsilon _{d}=0,$ $\tau $, $I$,
and $I_{s}$ are mirror symmetrical to $\phi =\pi $, and $\tau $ is always
negative. For positive and negative $\varepsilon _{d}$, $\tau $ and $I_{s}$
have just opposite properties: the peaks at $\phi =\pi /2$ (dips at $\phi
=3\pi /2$) for $\varepsilon _{d}=1$ correspond to the dips (peaks) for $%
\varepsilon _{d}=-1$ at the same $\phi $, but the curves for positive and
negative $\varepsilon _{d}$ intersect at $\phi =\pi $, as shown in Figs.
4(a) and (c). It hints us that by changing the gate voltage that is usually
utilized to alter the energy levels $\varepsilon _{d}$ in the QD, one can
adjust the STT. For example, when $\phi =\pi /2$, if we change $\varepsilon
_{d}$ from $-1$ to $1$, the STT changes from $0.06$ anticlockwise to $0.075$
clockwise. As the gate voltage is easier than the magnetic flux to control,
the present observation may offer a useful way to manipulate the magnetic
state of the free FM layer. The electrical current also displays quite
different oscillating behaviors for $\varepsilon _{d}=1$ and $-1$, which is
shown in Fig. 4(b).

Why can the STT be controlled by changing the magnetic flux and gate
voltage? Because the transmission probability of the spin-up electrons is
proportional to $cos(\theta +\phi +\gamma )$ and that of spin-down electrons
is proportional to $cos(\theta +\phi -\gamma )$ \cite{Sun1}. The spin-up and
spin-down electrons have different transmission probabilities if $\gamma $
is nonzero, leading to oscillations of the STT with magnetic flux $\phi $.
On the other hand, the STT is intimately related to the electrical current $I
$ \cite{Mu2}, and the magnitude of $I$ depends on the energy level $%
\varepsilon _{d}$ of QD, so it is reasonable that the STT can be manipulated
by adjusting the gate voltage. $I_{\uparrow }$ and $I_{\downarrow }$ give
rise to opposite STT on the right FM layer. Since $I_{\uparrow }$ and $%
I_{\downarrow }$ oscillate for various combination of $\phi $ and $\gamma $
in different ways, the STT may reach the maximum while $I$ is in its minimum
when $\phi $ and $\gamma $ take proper values.
\begin{figure}[tbp]
\includegraphics[width=0.6\linewidth,clip]{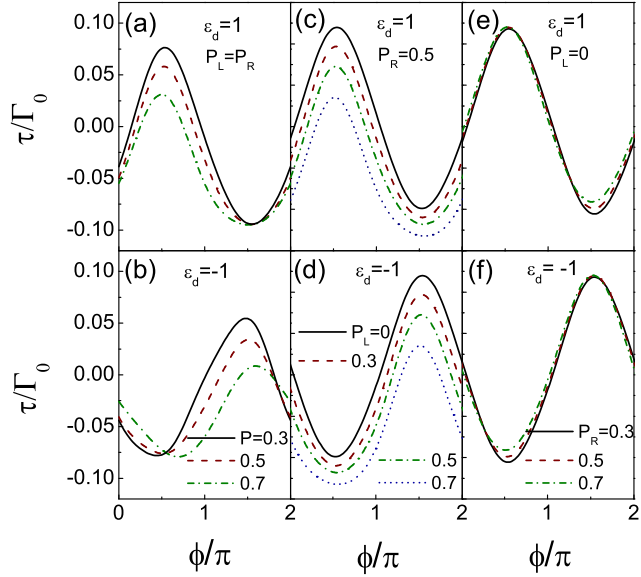}
\caption{(Color Online) The magnetic flux dependence of the spin transfer
torque for different spin polarizations at $\protect\theta =\protect\pi /3$,
and $\protect\gamma =\protect\pi /2$.}
\end{figure}

In addition, the spin polarization $P_{L,R}$ of the FM electrodes has also
effects on the magnetic flux dependence of the STT, as shown in Fig. 5.
Generally, with increasing $P_{L,R}$, the STT shows qualitatively similar
behaviors for positive and negative $\varepsilon _{d}$. When $P_{L}=P_{R}=P$%
, as shown in Figs. 5(a) and (b), the larger the polarization $P$, the
smaller the peaks of the STT. For different $P$, $\tau $ has obvious changes
when $0<\phi <3\pi /2$ for $\varepsilon _{d}=1$, and when $\pi /2<\phi <\pi $
for $\varepsilon _{d}=-1$. When $P_{R}$ and $P_{L}$ are different, e.g. $%
P_{R}=0.5$ and $P_{L}=0$, $0.3$, $0.7$, the larger $P_{L}$ is, the more
downward the curves move, as indicated in Figs. 5(c) and (d). It is
interesting that even the left electrode becomes spin unpolarized ($P_{L}=0$%
), the STT as a function of $\phi $ still behaves a sine-like curve and
retains almost intact for different spin polarizations $P_{R}$ [Figs. 5(e)
and (f)]. The existence of the STT at $P_{L}=0$ demonstrates that even if
the left electrode is a normal metal (NM), the unpolarized electrons from
the left NM lead flowing into the AB ring system with an QD encompassed by a
magnetic flux $\phi $ can become spin-polarized before entering into the
right FM electrode. It is apprehensible , because owing to the Rashba
effect, the spin-up and spin-down electrons pass through the AB ring system
at different transmission probabilities, as discussed above. When these
spin-polarized electrons flow into the right FM layer, they may transfer
some spin angular momenta to the local spins of the right FM electrode,
thereby giving rise to the STT. In this case, if $\phi =0$, the STT becomes
negligibly small. In the above analysis, we have presumed that the spin
relaxation time of electrons is greater than that of the tunneling time.
Thus, to ensure the feasibility of experimental observation, one must choose
proper materials as FM electrodes and QD, and design a viable ring system to
meet with the above requirements. It is interesting to note that a similar
mesoscopic ring system was proposed, where some material parameters were
discussed for possible experimental implementation \cite{radu} that may be
insightful for choosing proper materials for designing the present ring
system.

The effect of Rashba SO interaction $\gamma $ on the STT, electrical current
and spin current is shown in Fig. 6 for different magnetic flux $\phi $.
With increasing $\gamma $, when $\phi =0,$ the STT is always negative and
goes down non-monotonously. When $\phi =\pi /4$ or $\pi /2$, $\tau $ goes up
from negative to positive, reaches a round maximum, and then decreases, as
depicted in Figs. 6(a). This result implies that the STT can be enhanced
remarkably by matching $\phi $ with proper $\gamma $. The $\gamma $
dependences of the electrical current and spin current show different
behaviors for various $\phi $, as presented in Figs. 6(b) and (c). With
increasing $\gamma $, for $\phi =0$, both $I$ and $I_{s}$ increase; for $%
\phi =\pi /4$, $I$ first decreases to a round minimum, and then goes up,
while $I_{s}$ declines slowly; for $\phi =\pi /2$, the situation becomes
reverse, i.e., $I$ decreases dramatically, while $I_{s}$ first declines and
then goes up. In a word, the Rashba SO interactions have various effects on
the STT, electrical current and spin current.

\begin{figure}[tbp]
\includegraphics[width=0.6\linewidth,clip]{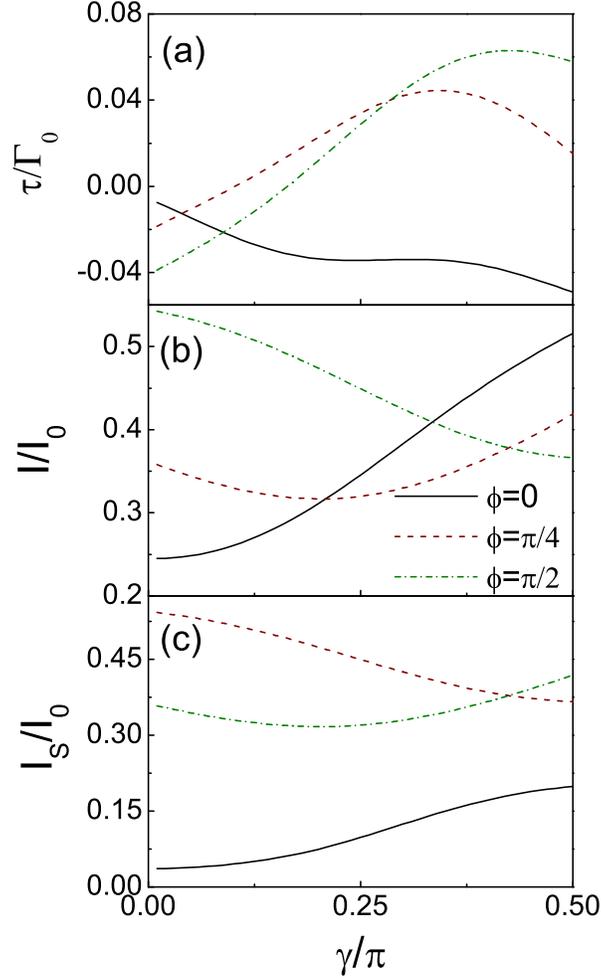}
\caption{(Color Online) The Rashba SO interaction $\protect\gamma $
dependence of (a) the spin transfer torque $\protect\tau $, (b) the
electrical current $I$, and (c) the spin current $I_{s}$ for different
magnetic flux $\protect\phi $, where $\protect\theta =\protect\pi /3$, and $%
\protect\varepsilon _{d}=1$.}
\end{figure}

Finally, the bias voltage dependences of the STT, electrical current and
spin current are studied for different $\gamma $ and $\phi $, as shown in
Fig. 7. In the simultaneous presence of $\gamma $ and $\phi $, e.g. $\gamma
=\phi =\pi /2$, with increasing the voltage, the STT first increases almost
linearly, reaches a peak, and then decreases slowly. After reaching zero, it
starts to increase again in a different direction. In the absence of either $%
\gamma $ or $\phi $ or both, $\tau $ is negative, and decreases
non-monotonously with increasing the bias voltage, as indicated in Fig.
7(a). For various combinations of $\gamma $ and $\phi $, the electrical
current $I$ exhibits qualitatively similar behaviors, which increases
overall in a non-ohmic way with increasing the bias [Fig. 7(b)]. For $\gamma
= \phi $, $I_{s}$ remains almost constant at a small bias. When the bias
passes a threshold, it increases linearly with the increase of $V$. For $%
\gamma \neq \phi $, $I_{s}$ grows up almost linearly despite of small
shoulders at a low bias, as displayed in Fig. 7(c). From Figs. 7(a) and (c),
we can see that the shoulder structure of the STT and the threshold of the
spin current appear around $eV=2\varepsilon _{d}$, where the resonant
tunneling happens. It is not surprising that the resonant tunneling has
influences on the spin-dependent transport of the system. However, it is
more important when $\gamma =\phi $, while it is negligible when $\gamma
\neq \phi $.

\begin{figure}[tbp]
\includegraphics[width=0.6\linewidth,clip]{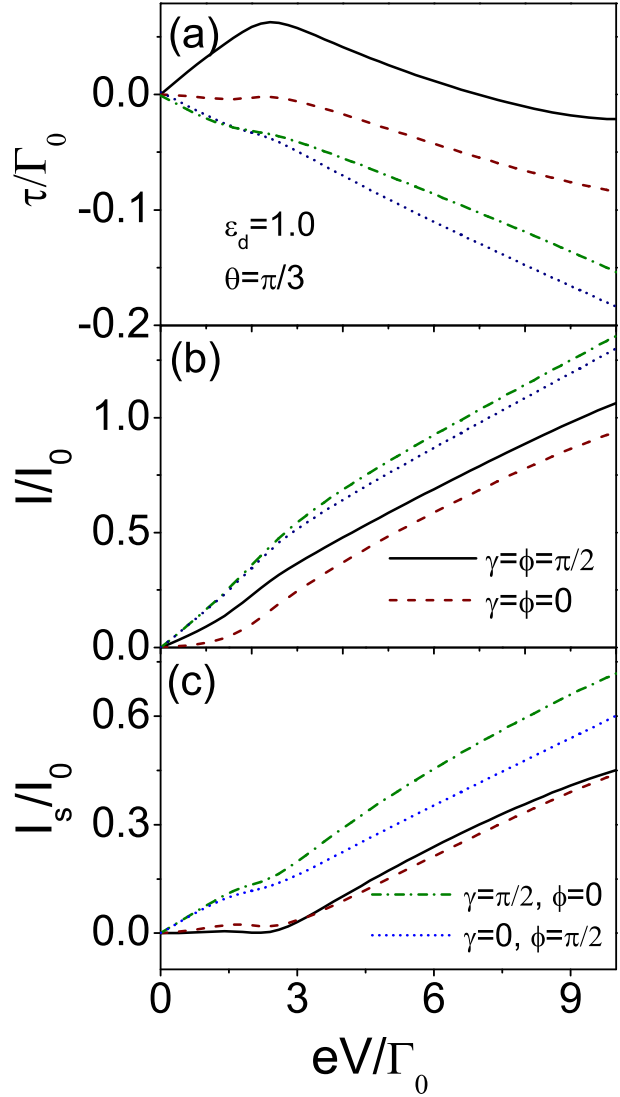}
\caption{(Color Online) The bias voltage dependence of (a) the spin transfer
torque $\protect\tau $, (b) the electrical current $I$, and (c) the spin
current $I_{s}$ for different $\protect\gamma $ and $\protect\phi $.}
\end{figure}

\section{Summary}

By means of the Keldysh nonequilibrium Green function method, we have
investigated the STE in the FM-QD(I)-FM ring system with Rashba SO
interactions. It has been found that both the direction and magnitude of the
STT are affected by the magnetic flux and the Rashba SO interactions. When
the SO interaction is strong enough, the STT acting on the spins of the
right FM electrode can be remarkably enhanced by matching the magnetic flux
through the AB ring, which makes it is possible to readily manipulate the
magnetic state of the free FM layer at a cost of lower electrical current.
This property is quite expected for nanospintronic devices where the
excessive heating generated by the electrical current should be avoided as
much as possible. It has also been uncovered that by adjusting the gate
voltage acting on the QD, both the magnitude and the direction the STT can
be changed, which gives an alternative way to manipulate the magnetic state
of the free FM layer. In addition, it is interesting to observe that the STT
can also be increased by the magnetic flux through the ring or the gate
voltage on the QD even if the left FM lead is changed to a NM.

We would like to mention that the results presented in this paper provide
useful information for designing practical spintronic devices based on the
STE. Such a ring layout can be used either as a memory element with a low
driving current or as a magnetometer to measure weak magnetic fields,
because the tunnel current depends sensitively on the magnetic flux threaded
the ring. On the other hand, the tunnel current or the magnetic state of the
free FM layer are affected by the Rashba SO interaction, and one may
inversely enable to estimate the magnitude of the Rashba SO interaction on
the QD by means of such a ring apparatus. We expect that the present
theoretical findings could be tested experimentally in future.

\acknowledgments

We are grateful to S. S. Gong, W. Li, X. L. Sheng, Z. C. Wang, Z. Xu, Q. B.
Yan, L. Z. Zhang and G. Q. Zhong for helpful discussions. This work is
supported in part by the National Science Fund for Distinguished Young
Scholars of China (Grant No. 10625419), NSFC (Grant Nos. 10934008,
90922033), the MOST of China (Grant No. 2006CB601102), and the Chinese
Academy of Sciences.


\begin{thebibliography}{99}
\bibitem{Berger} Berger L 1996 Phys. Rev. B \textbf{54} 9353

\bibitem{J.C} Slonczewski J C 1996 J. Magn. Magn. Mater. \textbf{159} L1

\bibitem{Zhu} Zhu Z G, Su G, Zheng Q R, and Jin B 2003 Phys. Rev. B \textbf{%
68} 224413; 2003 Phys. Lett. A \textbf{306} 249

\bibitem{Mu1} Mu H F, Zheng Q R, Jin B, and Su G 2005 Phys. Lett. A \textbf{%
336} 66

\bibitem{Mu2} Mu H F, Su G and Zheng Q R 2006 Phys. Rev. B \textbf{73} 054414

\bibitem{Ioannis} Theodonis I, Kioussis N, Kalitsov A, Chshiev M and Butler
W H 2006 Phys. Rev. Lett \textbf{97} 237205

\bibitem{Z.Z.Sun} Sun Z Z and Wang X R 2006 Phys. Rev. Lett \textbf{97}
077205

\bibitem{G.D} Fuchs G D, Katine J A, Kiselev S I, Mauri D, Wooley K S, Ralph
D C and Buhrman R A 2006 Phys. Rev. Lett \textbf{96} 186603

\bibitem{Levy} Levy P M and Fert A, Phys. Rev. Lett 2006 \textbf{97} 097205

\bibitem{J.A} Katine J A, Albert F J, and Buhrman R A 2000 Phys. Rev. Lett
\textbf{84} 3149

\bibitem{Z.Li} Li Z and Zhang S 2004 Phys. Rev. Lett \textbf{92} 207203

\bibitem{K.Xia} Xia K, Kelly P J, Bauer G E W, Brataas A, and Turek I 2002
Phys. Rev. B \textbf{65} 220401(R)

\bibitem{G.D2} Fuchs G D, Krivorotov I N, Braganca P M, Emley N C, Garcia A
G F, Ralph D C, and Buhrman R A 2005 Appl. Phys. Lett \textbf{86} 152509

\bibitem{S.Zhang} Zhang S, Levy P M, and Fert A 2000 Phys. Rev. Lett \textbf{%
88} 236601

\bibitem{M.D} Stiles M D and Zangwill A 2002 Phys. Rev. B \textbf{66} 014407

\bibitem{K.Ando} Ando K, Takahashi S, Harii K, Sasage K, Ieda J, Maekawa S,
and Saitoh E 2008 Phys. Rev. Lett \textbf{101} 036601

\bibitem{I.Z} Zutic I, Fabian J, and Sarma S D 2004 Rev. Mod. Phys \textbf{76%
} 323

\bibitem{Hanson} Harson R, Kouhenhoven L P, Petta J R, Tarucha S and
Vandersypen L M K 2007 \textbf{79} 1271

\bibitem{J.Mar} Martinek J, Utsumi Y, Imamura H, Barna\'{s} J, Maekawa S, K%
\"{o}nig J, and Sch\"{o}n G 2003 Phys. Rev. Lett \textbf{91} 127203

\bibitem{Xi} Chen X, Zheng Q R, and Su G 2007 Phys. Rev. B \textbf{76} 144409

\bibitem{Xi2} Chen X, Mu H F, Zheng Q R and Su G 2006 Phys. Lett. A \textbf{%
358} 47

\bibitem{Sun2} Sun Q F, Xie X C 2006 Phys. Rev. B \textbf{73} 235301

\bibitem{JPCM} Ying Y, Jin G and Ma Y 2009 J. Phys.: Condens. Matter \textbf{%
21} 275801

\bibitem{DD} Datta S and Das B 1990 Appl. Phys. Lett \textbf{56} 665

\bibitem{Sun1} Sun Q F, Wang J and Guo H 2005 Phys. Rev. B \textbf{71} 165310

\bibitem{X.W1} Waintal X and Brouwer P W 2001 Phys. Rev. B \textbf{63} 220407

\bibitem{X.W2} Waintal X and Brouwer P W 2002 Phys. Rev. B \textbf{65} 054407

\bibitem{Jauho} Haug H and Jauho A P \textit{Quantum Kinetics in Transport
and Optics of Semiconductors} (Springer, Berlin, 1998), p.166

\bibitem{J.Z} Sun J Z 2000 IBM J. Res. \& Dev. \textbf{50} 81

\bibitem{J.Z.Sun} Sun J Z 2000 Phys. Rev. B \textbf{62} 570

\bibitem{Xi3} Chen X, Zheng Q R, and Su G 2008 Phys. Rev. B \textbf{78},
104410

\bibitem{F.M} Mireles F and Kirzenow G 2001 Phys. Rev. B \textbf{68} 115316

\bibitem{T.M} Matsuyama T, Kursten R, Meissner C and Merkf U 2000 Phys. Rev.
B \textbf{61} 15588

\bibitem{D.G} Grundler D 2000 Phys. Rev. Lett \textbf{84} 6074

\bibitem{radu} Ionicioiu R and D'Amico I 2003 Phys. Rev. B \textbf{67}
041307(R)
\end{thebibliography}
\end{document}